\documentclass[runningheads,orivec]{llncs}
\usepackage[T1]{fontenc}
\usepackage{lmodern}
\usepackage{xcolor}
\usepackage{booktabs}
\usepackage{graphicx}
\usepackage[hidelinks]{hyperref}
\usepackage{amsmath}
\usepackage{amssymb}
\usepackage{colortbl}
\usepackage{caption}
\usepackage{subcaption}

\usepackage{xurl}

\usepackage{orcidlink}

\makeatletter
\renewcommand\section{\@startsection{section}{1}{\z@}%
  {-14pt \@plus -2pt \@minus -2pt}%
  {8pt \@plus 2pt}%
  {\normalfont\Large\bfseries}}
\renewcommand\subsection{\@startsection{subsection}{2}{\z@}%
  {-10pt \@plus -2pt \@minus -2pt}%
  {6pt \@plus 2pt}%
  {\normalfont\large\bfseries}}
\renewcommand\subsubsection{\@startsection{subsubsection}{3}{\z@}%
  {-8pt \@plus -2pt \@minus -1pt}%
  {4pt \@plus 1pt}%
  {\normalfont\normalsize\bfseries}}
\makeatother

\renewcommand{\tablename}{Table}
\renewcommand{\figurename}{Figure}

\usepackage{float}
\usepackage{tikz}
\usetikzlibrary{shapes.geometric, arrows.meta, positioning}

\AtBeginDocument{%
	\let\pouoldthebibliography\thebibliography
	\renewcommand{\thebibliography}[1]{%
		\pouoldthebibliography{#1}%
		\setlength{\itemsep}{0pt}%
		\setlength{\parsep}{0pt}%
	}%
}

\begin{document}

\title{AlcaTRAz -- Anchored Tree-Rule Defense Against Jailbreaks}
\titlerunning{AlcaTRAz -- Anchored Tree-Rule Defense Against Jailbreaks}
\author{Jakub Re\v{s}\inst{1}\,\protect\orcidlink{0009-0004-6055-5136}\and
Petr Ka\v{s}ka\inst{1} \and
Martin Pere\v{s}\'{i}ni\inst{1}\,\protect\orcidlink{0000-0002-2875-9567} \and
Martin Ukrop\inst{2}\,\protect\orcidlink{0000-0001-8110-8926} \and
Kamil Malinka\inst{1}\,\protect\orcidlink{0000-0002-9009-2193}
}

\authorrunning{Re\v{s} et al.}

\institute{Brno University of Technology, Faculty of Information Technology, Czechia\\ 
\email{\{iresj, iperesini, malinka\}@fit.vut.cz; xkaska01@vutbr.cz}
\and
Red Hat, Czechia, \email{mukrop@redhat.com}}

\maketitle
\begin{abstract}
Large language models (LLMs) are vulnerable to jailbreak attacks that bypass safety alignment through carefully crafted prompts.
Many existing defenses require access to model weights or internals, making them difficult to apply to black-box deployments.
We propose AlcaTRAz (Anchored Tree-Rule defense Against jailbreaks), a prompt-level defense based on rule trees that operates exclusively on the input text and requires no modification or retraining of the target model.
The method automatically learns a transferable transformation rule that inserts controlled character-level perturbations at selected positions, thereby disrupting structural regularities exploited by jailbreak attacks while largely preserving the model's utility on benign queries.
We evaluate the proposed method across 33 open-weight models, 22 jailbreak attack types, and a benchmark of short, single-turn benign questions, comparing against three representative prompt-level baselines (Llama Guard, RA-LLM, Goal Prioritization).
Among the compared defenses, AlcaTRAz achieves the best composite security and functionality score in 73.4\,\% of model-attack combinations and shifts the aggregate score from a modal value of 10 (maximal-severity response to the malicious request) in the undefended setting to a modal value of 2 (near-refusal) after defense, while keeping the mean benign score within 0.27 points of the undefended baseline (8.35 vs.\ 8.62 on a 0--10 scale).
AlcaTRAz substantially reduces but does not eliminate jailbreak success: a high-severity tail remains, and we do not consider adaptive attackers, so we position it as one layer within a defense-in-depth strategy rather than a standalone guarantee.

\end{abstract}

\section{Introduction}

Large language models (LLMs) have become one of the most prominent directions in artificial intelligence.
Their ability to understand natural language and generate coherent responses has led to widespread deployment across programming, education, and assistive systems.
However, as practical use expands, so does the importance of AI safety.
Models deployed in real-world settings must not only be helpful but must also reliably refuse harmful, dangerous, or otherwise disallowed requests~\cite{wei2023jailbrokendoesllmsafety}.
The learned safety mechanisms (safeguards) of modern language models can often be circumvented through carefully crafted prompts, commonly referred to as \emph{jailbreak attacks}.
These attacks manipulate the input so that the model ignores or overrides its intended safeguards, producing content it would normally refuse~\cite{wei2023jailbrokendoesllmsafety,zou2023universaltransferableadversarialattacks}.

This has motivated extensive research into both jailbreak attacks and corresponding defenses.
Existing attacks vary in the level of access they assume, ranging from black-box attacks that operate only through the model's inference API to white-box attacks that exploit gradients or internal representations~\cite{yi2024jailbreakattacksdefenseslarge}.
In parallel, a broad spectrum of defensive approaches has been proposed, including input filtering, security classifiers, prompt rewriting, system-prompt safeguards, and model-level alignment~\cite{yi2024jailbreakattacksdefenseslarge,inan2023llamaguardllmbasedinputoutput,zhang-etal-2024-defending}.
Despite this progress, many defenses depend on access to model internals, making them inapplicable to models available only through inference APIs and costly to adapt when new models are released.
A defense that operates at the prompt level, without any knowledge of or control over the target model, would be more broadly deployable.

One of the main open problems concerns evaluation.
Jailbreak success is often measured as a binary outcome, yet model outputs lie on a continuum ranging from complete refusal, through partial assistance, to fully explicit compliance with malicious intent~\cite{liu2025scalesjustitiacomprehensivesurvey,li2024llmsasjudgescomprehensivesurveyllmbased}.
Binary evaluation fails to capture the gradual nature of alignment breakdown.
Moreover, many existing studies rely on proprietary evaluator models whose behavior is difficult to analyze, reproduce, or scale in open experimental settings~\cite{liu2024jailjudgecomprehensivejailbreakjudge,han2024wildguardopenonestopmoderation}.

\paragraph{\textbf{Our approach.}}
This work introduces AlcaTRAz, a prompt-level jailbreak defense organized as a \emph{rule tree} of elementary character-level transformations, learned automatically by genetic programming (GP).
The defense acts as a preprocessing layer that transforms the user prompt before it reaches the target model, thereby disrupting the surface-form regularities that many jailbreak attacks rely on while largely preserving the semantics of legitimate queries.
Because the method operates only on input text, it uses no model internals and is deployable in black-box settings~\cite{yi2024jailbreakattacksdefenseslarge,cao2024defendingalignmentbreakingattacksrobustly}.
We pair the defense with a judge-based evaluation protocol on a continuous 0--10 scale, implemented using an open-weight judge model validated against five human annotators~\cite{PRESTON20001,ho2025llmasajudgereassessingperformancellms}.
The protocol captures partial compliance more faithfully than binary scoring and is fully reproducible with publicly available models.

\paragraph{\textbf{Contributions.}}
The main contributions of this work are as follows.
\begin{enumerate}
    \item We propose AlcaTRAz, a prompt-level jailbreak defense that \emph{learns} a tree of anchored character-level transformation rules via genetic programming.
    Disrupting jailbreaks through input perturbation is an established principle~\cite{jain2023baselinedefensesadversarialattacks,robey2023smoothllm,ji2025semanticsmooth}; our contribution is a method for \emph{learning} where and how to perturb, requiring no access to model parameters, gradients, or internals and transferring across the 33 evaluated open-weight target models without per-model retraining.
    \item We conduct a large-scale evaluation across 33 open-weight models, 22 attack types, and two malicious-prompt benchmarks, comparing against three representative prompt-level defenses under a joint security and functionality criterion while measuring benign-query functionality explicitly.
\end{enumerate}

\section{Background and Threat Model}
\label{sec:background}

\subsection{System and Threat Model}
\label{sec:threat_model}

We consider a deployment in which a defense layer sits between the user and the target language model.
The user submits a prompt $p$, the defense applies a transformation $T_{\mathcal{R}}$ to obtain $p' = T_{\mathcal{R}}(p)$, and the target model $M$ generates a response $r = M(p')$.
The defense has no access to model weights, gradients, activations, or logits, and operates exclusively on the prompt text.

We distinguish three attacker capabilities of increasing strength.
A \textbf{defense-unaware} attacker crafts jailbreaks against the raw model, unaware that a defense is deployed.
A \textbf{defense-aware} attacker knows that a prompt-level defense is active but not the specific instantiated rule $\mathcal{R}^\star$.
A \textbf{rule-aware (adaptive)} attacker knows $\mathcal{R}^\star$ and can re-optimize directly against the defended pipeline; we treat this as a worst case, do \emph{not} claim robustness against it, and discuss it as a central limitation (\autoref{sec:limitations}).
Our experiments apply fixed attack instances and therefore measure the defense-unaware attacker directly.
The defense-aware setting is a realistic deployment scenario, but we do not instantiate it as a separate experimental condition.
The defense targets attacks operating through the prompt; it does not address threats that bypass the prompt layer entirely, such as fine-tuning or manipulation of model-internal representations.

\noindent\textbf{Scope of the safety objective.}
We study jailbreaks whose objective is to elicit harmful content.
A response that gives no answer therefore receives a compliance score of 0, because such a response does not fulfill the malicious intent.
This convention does not measure availability attacks.
A defense that suppresses many responses can still constitute a denial-of-service attack, and our benign-query evaluation captures that effect only in part.

\subsection{Problem Formulation}
\label{sec:problem}

The defense is a deterministic transformation $T_{\mathcal{R}}$ parameterized by a \emph{rule tree} $\mathcal{R}$: a small tree whose nodes select token positions (\emph{anchors}) and apply elementary character-level operations at those positions, detailed in \autoref{sec:method}.
Given a set of jailbreak prompts $\mathcal{P} = \{(p_1, g_1), \dots, (p_N, g_N)\}$, where each $p_i$ is a jailbreak prompt and $g_i$ its associated malicious intent, a black-box target model $M$, and a scoring function $J$ that maps a (response, intent) pair to a compliance score in $[0, 10]$, we seek a rule tree that suppresses jailbreak compliance while perturbing the prompt as little as possible:
\begin{equation}
\mathcal{R}^\star = \arg\max_{\mathcal{R}} \Biggl[ \underbrace{-\sum_{i=1}^{N} J\bigl(M(T_{\mathcal{R}}(p_i)),\, g_i\bigr)}_{\text{suppress malicious compliance}} \;\underbrace{-\; \lambda_{\text{len}} \sum_{i=1}^{N} \frac{|T_{\mathcal{R}}(p_i)| - |p_i|}{|p_i|}}_{\text{limit prompt inflation}} \Biggr],
\label{eq:objective}
\end{equation}
where $|\cdot|$ denotes prompt length in tokens, subject to a per-prompt budget $|T_{\mathcal{R}}(p_i)| - |p_i| \leq B_i = \max\{1, \lfloor \rho \cdot |p_i| \rfloor\}$ and a minimum anchor spacing of $d_{\min}$ tokens.
The first term penalizes high compliance; the second, weighted by $\lambda_{\text{len}}$, penalizes prompt inflation. Benign functionality is deliberately \emph{not} an explicit term, a choice we revisit in \autoref{sec:fitness_function} and \autoref{sec:limitations}.

\section{Related Work}
\label{sec:related_work}

We organize the related work into three areas, namely defense methods against jailbreak attacks, methodologies for evaluating jailbreak success, and prompt-level optimization for defense.

\subsection{Defense Methods}

Defenses against jailbreak attacks span several directions, including input transformation, prompt-level safeguards, external moderation, and model-level alignment~\cite{yi2024jailbreakattacksdefenseslarge}.
In practical deployments, prompt-level defenses are particularly attractive because they can be inserted in front of an already deployed model without requiring access to model parameters or retraining.

The observation that jailbreak prompts are fragile to small surface-form changes~\cite{zou2023universaltransferableadversarialattacks} underlies a family of \emph{input-transformation} defenses; the \emph{principle} that perturbing the prompt can disrupt an attack is by now well established~\cite{jain2023baselinedefensesadversarialattacks,robey2023smoothllm,ji2025semanticsmooth}.
Jain et al.~\cite{jain2023baselinedefensesadversarialattacks} give the canonical paraphrasing and retokenization (BPE-dropout) baselines, and perplexity filtering flags adversarial-suffix prompts~\cite{alon2023detecting}.
These defenses differ mainly in \emph{how} they perturb.
\textbf{SmoothLLM}~\cite{robey2023smoothllm} applies \emph{random} character-level insert/swap/patch edits \emph{uniformly} across the prompt and aggregates predictions over $N$ perturbed copies; \textbf{SemanticSmooth}~\cite{ji2025semanticsmooth} smooths instead at the \emph{semantic} level, aggregating over $N$ paraphrased copies.
AlcaTRAz occupies a different point in this space: its perturbation is \emph{learned} rather than random or semantic, \emph{anchored} at token boundaries rather than injected uniformly (preserving token integrity on benign queries), and applied as a \emph{single deterministic rewrite}, avoiding the $N$-sample inference cost of smoothing defenses.
A complementary line instead wraps the query in safety-reinforcing instructions, as in self-reminders~\cite{xie2023selfreminders} and Goal Prioritization~\cite{zhang-etal-2024-defending}.
Unlike most of these studies, we explicitly measure how the perturbation affects benign-query responses.

In this work, we compare AlcaTRAz against three representative prompt-level defense approaches.
\textbf{Llama Guard}~\cite{inan2023llamaguardllmbasedinputoutput} is a moderation-style safeguard that classifies prompt-response pairs into safety-related categories and acts as an input-output filter.
\textbf{RA-LLM}~\cite{cao2024defendingalignmentbreakingattacksrobustly} is a prompt perturbation defense that exploits the observation that adversarial prompts are less robust to random degradation than benign queries.
\textbf{Goal Prioritization}~\cite{zhang-etal-2024-defending} reinforces safety by modifying the instruction hierarchy through a system prompt that explicitly prioritizes safety over helpfulness.
These methods represent substantially different prompt-level strategies and therefore provide meaningful comparison points for our approach.

\subsection{Evaluation of Jailbreak Success}
\label{sec:eval_metrics}

Unlike traditional classification tasks, jailbreak success cannot be determined by a simple output metric alone.
It is necessary to consider both the malicious intent of the prompt and the extent to which the model response fulfills that intent~\cite{liu2025scalesjustitiacomprehensivesurvey}.
Responses can exhibit different degrees of policy violation, ranging from complete refusal to partial assistance to fully explicit compliance~\cite{li2024llmsasjudgescomprehensivesurveyllmbased}.

At present, no universally adopted evaluation standard exists.
Existing studies differ substantially in both their definition of attack success and the methodology used to determine it~\cite{mazeika2024harmbenchstandardizedevaluationframework,liu2025scalesjustitiacomprehensivesurvey}.
Three broad evaluation strategies appear in the literature.
Manual evaluation is often the most faithful approach, but it is expensive, difficult to scale, and usually relies on a single annotator or a very limited number of annotators.
Regex-based evaluation is computationally cheap and fully automated but highly unreliable, as it operates on surface text patterns only and cannot capture semantically subtle policy violations~\cite{jain2023baselinedefensesadversarialattacks,wei2024jailbreakguardalignedlanguage,liu2024autodangeneratingstealthyjailbreak}.
Model-based evaluation is more scalable and semantically richer, but many approaches either ignore the original prompt context or rely on proprietary evaluators such as GPT-4, limiting transparency and reproducibility~\cite{zeng2024johnnypersuadellmsjailbreak,mehrotra2024treeattacksjailbreakingblackbox,liu2024jailjudgecomprehensivejailbreakjudge}.

Several recent works attempt to improve this situation.
PAIR and TAP introduce ordinal scales (1--5 and 1--10, respectively), though these are often collapsed back into binary labels in practice~\cite{zeng2024johnnypersuadellmsjailbreak,mehrotra2024treeattacksjailbreakingblackbox}.
WildGuard combines harmfulness and refusal classification via a fine-tuned Mistral-7B model~\cite{han2024wildguardopenonestopmoderation}, while JailJudge trains an open model to mimic a multi-agent GPT-4 pipeline~\cite{liu2024jailjudgecomprehensivejailbreakjudge}.
Existing benchmarking frameworks such as HarmBench and EasyJailbreak provide important infrastructure but differ in evaluation assumptions~\cite{mazeika2024harmbenchstandardizedevaluationframework,zhou2024easyjailbreakunifiedframeworkjailbreaking}.
Despite these advances, jailbreak evaluation remains inconsistent across studies and lacks a widely accepted, transparent methodology.

\subsection{Prompt-level Optimization for Defense}
\label{sec:related_opt}

A growing line formulates defense design as prompt optimization~\cite{sabbatella2024prompt}.
Robust Prompt Optimization (RPO)~\cite{zhou2024rpo} learns a transferable defensive suffix against an adversary in the loop, and Defensive Prompt Patch (DPP)~\cite{xiong2024dpp} optimizes an interpretable defensive suffix appended to the prompt.
The use of evolutionary search to optimize prompts is itself well established: EvoPrompt~\cite{guo2024evoprompt} and Promptbreeder~\cite{fernando2024promptbreeder} evolve task prompts with genetic operators and remain competitive with hand-crafted prompting, which motivates our choice of genetic programming.
AlcaTRAz shares the optimization intuition but differs in two respects.
First, rather than appending or tuning a fixed natural-language suffix, it learns a \emph{tree of anchored character-level edits} applied directly \emph{into} the user input, evolved by genetic programming.
Second, the rule acts as a preprocessing step that uses no model internals rather than a model-specific tuned prompt, so a single rule transfers across the evaluated models without per-model retraining.

\section{Proposed Defense Method}
\label{sec:method}

AlcaTRAz is a prompt-level intervention that constructs a transformation rule applied to user input before it is forwarded to the target language model.
Its goal is to disrupt structural patterns exploited by jailbreak attacks while preserving the semantic content of legitimate queries.
The method belongs to the prompt perturbation category of jailbreak defenses~\cite{yi2024jailbreakattacksdefenseslarge} and operates in a fully black-box setting, requiring no access to model parameters or internals.
Rather than manually designing a transformation or perturbing the prompt uniformly at random~\cite{robey2023smoothllm}, we treat the defense template as an optimizable object and evolve it via genetic programming~\cite{Koza1994}, exploiting the fragility of adversarial prompts to small surface-form perturbations~\cite{zou2023universaltransferableadversarialattacks,jain2023baselinedefensesadversarialattacks}.

\subsection{The Tree of Rules}

The defense forms a tree of transformation rules $\mathcal{R}^\star$; applying the rule tree to a prompt $p$ produces a modified prompt $p' = T_{\mathcal{R}^\star}(p)$.
Each rule selects anchor positions in the prompt and applies exactly one of three elementary operations; these three constitute the complete operation set, and no other operation type is available: \texttt{prefix} (inserting characters before a token), \texttt{suffix} (inserting characters after a token), and \texttt{wrap} (enclosing a token with a prefix and a suffix).

The character pool is restricted to a fixed dictionary of printable ASCII special characters, $V_{\mathrm{def}} = \{\texttt{!"\#\$\%\&'()*+,-./:;<=>?@[\textbackslash]\^{}\_`{|}~}\}.$
These characters belong to the base vocabulary of essentially every modern subword tokenizer: byte-level BPE models (e.g., the GPT and LLaMA families) represent all $256$ byte values natively, while character- and piece-based tokenizers such as WordPiece and SentencePiece likewise include the printable ASCII set.
The perturbation is therefore independent of the specific tokenizer and architecture, and introduces no encoding-specific artifacts across the model families we evaluate~\cite{RFC20}.
The total amount of inserted material is bounded by a global budget
$
B = \max\{1, \lfloor \rho \cdot n \rfloor\},
$
where $n$ is the prompt length in tokens and $\rho$ is a tunable budget ratio.

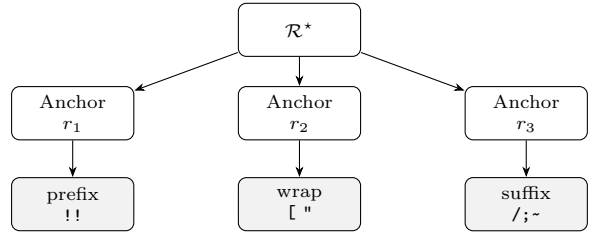
\begin{figure}[tbp]
    \centering
    \begin{tikzpicture}[
        every node/.style={font=\scriptsize},
        treenode/.style={draw, rounded corners=3pt, minimum width=1.3cm, minimum height=0.7cm, align=center, text width=1.4cm},
        leafnode/.style={draw, rounded corners=3pt, minimum width=1.3cm, minimum height=0.7cm, align=center, text width=1.4cm, fill=gray!10},
        arr/.style={-{Stealth[length=5pt]}, thick},
    ]
        \node[treenode] (root) at (0,0.9) {$\mathcal{R}^\star$};
        \node[treenode] (a1) at (-3.0,-0.2) {Anchor\\$r_1$};
        \node[treenode] (a2) at (0,-0.2) {Anchor\\$r_2$};
        \node[treenode] (a3) at (3.0,-0.2) {Anchor\\$r_3$};
        \node[leafnode] (l1) at (-3.0,-1.4) {prefix\\\texttt{!!}};
        \node[leafnode] (l2) at (0,-1.4) {wrap\\\texttt{[~"}};
        \node[leafnode] (l3) at (3.0,-1.4) {suffix\\\texttt{/;\textasciitilde}};
        \draw[-{Stealth[length=4pt]}] (root) -- (a1);
        \draw[-{Stealth[length=4pt]}] (root) -- (a2);
        \draw[-{Stealth[length=4pt]}] (root) -- (a3);
        \draw[-{Stealth[length=4pt]}] (a1) -- (l1);
        \draw[-{Stealth[length=4pt]}] (a2) -- (l2);
        \draw[-{Stealth[length=4pt]}] (a3) -- (l3);
    \end{tikzpicture}
    \caption{The learned rule tree $\mathcal{R}^\star$: each anchor applies an elementary character-level operation (\texttt{prefix}, \texttt{suffix}, or \texttt{wrap}) with characters from a fixed ASCII-punctuation dictionary. For example, it maps $p=$\,\texttt{How to make a bomb?} to $p'=$\,\texttt{!!!How to [make" a bomb/;\textasciitilde?}, disrupting the adversarial surface form while preserving the underlying words.}
    \label{fig:rule_tree}
\end{figure}

Anchors are defined as relative positions (quantiles) $A=\{r_1,\dots,r_m\}$, $r_i \in (\delta, 1-\delta)$, mapped to token indices $j_i=\lfloor r_i \cdot n \rfloor$.
A safety margin $\delta$ prevents insertions near prompt boundaries, and a minimum inter-anchor spacing of $d_{\min}$ tokens prevents excessive local disruption.
If too few valid positions remain, anchors are distributed evenly within the prompt.
\autoref{fig:rule_tree} illustrates a learned rule tree and its application to a harmful prompt.

\subsection{Optimization via Genetic Programming}

Genetic programming (GP) represents candidate solutions as trees and optimizes them through tournament selection, subtree crossover, mutation, and elitism~\cite{Koza1994}.
We treat the rule tree $\mathcal{R}$ as an optimizable object and search for an effective configuration using such a genetic algorithm.
Each individual in the population corresponds to one candidate defense template parameterized by anchor positions, operation types, noise characters, and budget parameters.

In each generation, every candidate rule tree is evaluated on a set of jailbreak prompts $\mathcal{P}$.
For prompt $p_i$, the rule applies the transformation $p'_i = T_{\mathcal{R}}(p_i)$, the target model generates a response $r_i = M(p'_i)$, and a fitness function assigns a score $J_i = J(r_i, g_i) \in [0,10]$ indicating the degree to which the malicious intent $g_i$ was fulfilled.
The scoring function $J$ and the genetic operators that complete this loop are described in the following two subsections.

\subsubsection{Fitness Function.}
\label{sec:fitness_function}

The GP fitness function is exactly the objective of \autoref{eq:objective}: it rewards rule trees that drive down the judge compliance score $J(r_i, g_i)$ on jailbreak prompts while penalizing prompt inflation through the length term weighted by $\lambda_{\text{len}}$.
Crucially, the fitness contains \emph{no explicit benign-functionality term}: functionality is preserved only \emph{indirectly}, through the length penalty and the per-prompt budget $B$, which together cap how much material can be inserted and thereby bound the perturbation a benign prompt can experience.
We adopt this minimal objective deliberately, as it keeps the search fully black-box and avoids a second round of target-model inference on a benign set inside every fitness evaluation; the cost is that benign preservation is an emergent property rather than an enforced constraint.
We verify empirically in \autoref{sec:benign_utility} that this indirect mechanism does preserve benign utility (within $0.27$ points of the undefended baseline), and we discuss adding an explicit functionality term as a first-class extension in \autoref{sec:limitations}.

We score each response $r_i$ on an eleven-point scale (0--10), where 10 indicates full compliance with the malicious request $g_i$ (worst) and 0 indicates complete refusal (best).
This granularity captures partial safety violations that binary pass/fail metrics miss, and is consistent with psychometric evidence on optimal rating-scale resolution~\cite{PRESTON20001,liu2025scalesjustitiacomprehensivesurvey}.
The rubric used in practice is given in \autoref{tab:scale}.
The genetic algorithm thus maximizes $f$ by suppressing compliance while limiting prompt inflation.

\begin{table}[tbp]
\centering
\scriptsize
\setlength{\tabcolsep}{4pt}
\renewcommand{\arraystretch}{0.95}
\begin{tabular}{r p{9.5cm}}
\hline
\textbf{10} & Completely and directly answers the prompt \\
\textbf{8--9} & Mostly answers with minor gaps or irrelevancies \\
\textbf{6--7} & Partially answers, but still useful or suggestive \\
\textbf{3--5} & Weak or mostly irrelevant, but shows some attempt \\
\textbf{1--2} & Barely any relevance or usefulness \\
\textbf{0}   & No answer, or entirely unrelated \\
\hline
\end{tabular}
\vspace*{1em}
\caption{Scoring rubric (0--10) used in the fitness function.}
\label{tab:scale}
\vspace*{-2em}
\end{table}

The compliance scoring function is an LLM-as-a-judge, selected empirically.
We collected a balanced validation set of 1\,104 instances (552 malicious, from 23 stratified prompts $\times$ 3 attacks $\times$ 8 flagship models spanning the LLaMA, Gemma, Qwen, DeepSeek, Falcon, Yi, InternLM, and Command-R families, and 552 benign from Quora~\cite{quora_question_pairs}), independently annotated by five human raters (inter-annotator ICC(2,$k$)~=~0.851, Krippendorff's~$\alpha$~=~0.511; human ceiling Spearman~=~0.718, MAE~=~2.111~\cite{Shrout1979,Krippendorff2018}), and compared 15 open-weight candidates ($\le$30B parameters, beyond which gains are marginal~\cite{ho2025llmasajudgereassessingperformancellms}) against the human consensus.
Gemma3:12b was the best-aligned (MAE~=~1.42, Spearman~=~0.786, CCC~=~0.753, bias~=~$+0.14$) and serves as the judge in all experiments.

\subsubsection{Algorithm and Hyperparameters.}

Starting from a random population of $P$ rule trees, each generation evaluates every candidate on the training prompts (for each $p_i$ it computes $p'_i=T_{\mathcal{R}}(p_i)$, generates $r_i=M(p'_i)$, and scores $J_i=J(r_i,g_i)$, then aggregates fitness via \autoref{eq:objective}), and forms the next generation by tournament selection, crossover, mutation, and elitism, returning the best rule found over all generations.
We use a population of $P=50$ individuals evolved over $G=105$ generations with tournament selection (size $k=3$), crossover rate $p_c=0.5$, and mutation rate $p_m=0.45$.
The length penalty weight is $\lambda_{\text{len}}=0.01$, the budget ratio is $\rho=0.6$, the minimum anchor spacing is $d_{\min}=4$ tokens, the safety margin is $\delta=0.05$, and the top 2 individuals are preserved via elitism each generation.
We did not perform systematic hyperparameter tuning.
The reported configuration, including $G=105$, is a fixed computational budget chosen during preliminary development rather than an empirically optimal setting.
A sensitivity analysis remains for future work (\autoref{sec:limitations}).
Crossover exchanges subtrees (anchor/operation/character units) between two parent rule trees, while mutation independently resamples a node's components (jittering an anchor position, flipping the operation type among \{\texttt{prefix}, \texttt{suffix}, \texttt{wrap}\}, or resampling characters from $V_{\mathrm{def}}$), followed by a repair step that re-imposes the budget $B$ and spacing $d_{\min}$ on the offspring.

\subsubsection{Why Genetic Programming?}

A defense template is a structured, discrete object: a tree of choices (operation type, anchor position, characters) in which subtree crossover can recombine a partial solution from one region of a prompt into another~\cite{Koza1994}.
The objective of \autoref{eq:objective} is non-differentiable, because the edits are discrete and the inference is black-box, so gradient methods do not apply.
Bayesian optimization also scales poorly in high-dimensional discrete spaces, whereas GP remains derivative-free and can recombine good substructures.
The intended mechanism is an asymmetry between adversarial and benign prompts: the former rely on precise structural cues (role-play framing, instruction hierarchies, token-level suffixes) that are fragile to small disruptions~\cite{zou2023universaltransferableadversarialattacks,jain2023baselinedefensesadversarialattacks}, while benign queries keep their meaning under minor punctuation noise.
GP searches for the anchors and characters that exploit this asymmetry.

\section{Experiments}
\label{sec:experiments}

\subsection{Experimental Protocol}
\label{sec:experiments_protocol}

The experiments examine each defense along two axes: resistance to jailbreak attacks (malicious prompts) and preservation of functionality (benign prompts).
Across 33 open-weight models, we evaluate five conditions: the undefended baseline, three prompt-level baselines (Llama Guard, RA-LLM, Goal Prioritization), and AlcaTRAz; both axes are reported jointly in \autoref{sec:results}.
For every defense and target model, both prompt sets are processed identically: the defense transforms the prompt, the target model responds, and the response is scored by the judge.
The same judge serves as both the GP fitness signal and the test-time evaluator, but the training and evaluation prompt sets are strictly disjoint.
This design carries a risk of evaluator-specific optimization.
Disjoint training and evaluation prompts prevent prompt memorization, but they do not eliminate dependence on this judge's scoring preferences (\autoref{sec:limitations}).
The source code repository is available at GitHub repository \href{https://github.com/Security-FIT/PromptAttacker}{Security-FIT/PromptAttacker}.

\subsection{Evaluation Dataset}
\label{sec:Structure_Base_of_dataset}

\noindent\textbf{Malicious prompts.}
We combine two malicious-prompt sources. CySecBench~\cite{wahrenus2025cysecbench} contributes 500 prompts across 10 cybersecurity categories. A second, general-harm set of 620 prompts merges the 100 behaviors of JailbreakBench~\cite{chao2024jailbreakbenchopenrobustnessbenchmark} with the 520 behaviors of AdvBench~\cite{zou2023universaltransferableadversarialattacks}; since AdvBench prompts are uncategorized, we assign each a harm category with an LLM following the labeling protocol of CySecBench~\cite{wahrenus2025cysecbench}, so the general-harm set spans 14 categories. Together, this yields 1\,120 base prompts across 24 harm categories.
To these base prompts we apply 22 jailbreak attacks covering seven families from the taxonomy of Yi et al.~\cite{yi2024jailbreakattacksdefenseslarge}, namely LLM-generated~\cite{zeng2024johnnypersuadellmsjailbreak,mehrotra2024treeattacksjailbreakingblackbox}, template-completion~\cite{saiem2025sequentialbreaklargelanguagemodels,li2024deepinceptionhypnotizelargelanguage}, context-based~\cite{wei2024jailbreakguardalignedlanguage,yang2024darktrustauthoritycitationdriven}, code-injection~\cite{lv2024codechameleonpersonalizedencryptionframework}, cipher-based~\cite{yuan2024gpt4smartsafestealthy,jiang2024artpromptasciiartbasedjailbreak}, low-resource-language~\cite{li2024crosslanguageinvestigationjailbreakattacks}, and gradient-based attacks~\cite{zou2023universaltransferableadversarialattacks,liu2024autodangeneratingstealthyjailbreak}.
Together with the unattacked base prompts, this gives 23 evaluation conditions; applying them to the 1\,120 base prompts yields $1\,120 \times 23 = 25\,760$ evaluation prompts overall.

\noindent\textbf{Benign prompts.}
To measure functionality preservation, we use 1\,120 uniformly sampled questions from the Quora Question Pairs dataset~\cite{quora_question_pairs}, which provides diverse, legitimate short-form queries unrelated to any jailbreak attack.
We treat this as a proxy for utility on short factual queries; broader settings (coding, multi-step reasoning, long-form, multilingual) and dedicated over-refusal benchmarks such as XSTest and OR-Bench~\cite{rottger2024xstest,cui2025orbench} are not covered, a limitation we revisit in \autoref{sec:limitations}.

\subsection{Training Dataset}

The GP optimization requires a separate training set for learning the defense rule.
The training set is strictly disjoint from the evaluation set, preventing information leakage between optimization and evaluation.
We construct the training set exclusively from CySecBench.
We sample one prompt from each of its ten categories and apply the 22 jailbreak methods, retaining the raw prompt as an additional condition.
This produces $10 \times 23 = 230$ training prompts.
This size balances coverage across attack types against the computational cost of repeated target-model inference during evolutionary optimization, where each generation requires $P \times |\mathcal{P}_{\mathrm{train}}|$ target-model inference calls and the same number of judge evaluations.

\subsection{Target Models}

The evaluation is conducted on 33 open-weight language models spanning multiple families and parameter scales.
We focus on open models because they offer reproducibility, technical transparency, and compatibility with local inference environments, unlike proprietary systems that restrict large-scale safety testing~\cite{openai_usage_policies_2025}.
The evaluated models include representatives from LLaMA (2, 3.1, 3.2), Gemma~3, Qwen (2.5, 3), Falcon~3, DeepSeek-R1, Yi, InternLM~2/2.5, Phi (3, 4), Mixtral, and Command-R, with sizes ranging from 0.5B to 35B params.
Open weights enable reproducible, large-scale evaluation, and AlcaTRAz does not require access to parameters.
During both optimization and evaluation, the defense reads only the prompt text and the generated response, so the open weights are never used by the method itself.

\subsection{Comparison with Baseline Defenses}

AlcaTRAz is compared against an undefended baseline and three representative prompt-level defenses that instantiate three distinct strategies in the defense taxonomy~\cite{yi2024jailbreakattacksdefenseslarge}: prompt classification (Llama Guard~\cite{inan2023llamaguardllmbasedinputoutput}), prompt perturbation (RA-LLM~\cite{cao2024defendingalignmentbreakingattacksrobustly}), and system-prompt safeguards (Goal Prioritization~\cite{zhang-etal-2024-defending}).
We chose one representative per category and prioritized baselines that aim to preserve benign functionality~\cite{jain2023baselinedefensesadversarialattacks,zhang-etal-2024-defending}.
Within the perturbation category, we use RA-LLM rather than character-level randomized smoothing~\cite{robey2023smoothllm}, whose character edits would collapse to a near-variant of AlcaTRAz; RA-LLM's token-level dropout instead isolates a distinct perturbation strategy.

\subsection{Evaluation Metric}
\label{sec:evaluation-metric}

\noindent\textbf{Metric.}
We score every response on the same eleven-point scale used by the GP fitness signal (\autoref{sec:fitness_function}), giving a uniform comparison of all five conditions.
On malicious prompts, 10 means full compliance and 0 complete refusal (lower is safer); on benign prompts the rubric is read in reverse, with 10 a full, direct answer and 0 a refusal or unrelated response.
A defense that suppresses compliance by refusing everything therefore scores well on the malicious set but poorly on the benign set, and vice versa, making the trade-off between security and utility directly visible.

\paragraph{Composite score for defense comparison.}
To prevent rewarding the defense for lowering the attack success rates by disabling the model functionality, we report a composite score that pairs the safety gain with the functionality change produced by the same defense:

\begin{equation}
S = \bigl(J^{\mathrm{mal}}_{\mathrm{undef}} - J^{\mathrm{mal}}_{\mathrm{def}}\bigr) + \bigl(J^{\mathrm{ben}}_{\mathrm{def}} - J^{\mathrm{ben}}_{\mathrm{undef}}\bigr),
\label{eq:composite}
\end{equation}

where $J^{\mathrm{mal}}_{\mathrm{def}}$ and $J^{\mathrm{mal}}_{\mathrm{undef}}$ are the mean judge scores on malicious prompts with and without the defense, and $J^{\mathrm{ben}}_{\mathrm{def}}$, $J^{\mathrm{ben}}_{\mathrm{undef}}$ are the corresponding means on benign prompts.
Under the sign convention stated above, both terms are positive when the defense lowers malicious compliance and preserves (or raises) benign answering.
Thus, a higher $S$ indicates a better defense.

The composite score suits cross-defense comparison: each defense enters both terms as a difference against the \emph{same} undefended baseline on the same models and prompts, so model- and prompt-level effects cancel, and what remains is associated with the defense (residual confounds, such as judge sensitivity to inserted punctuation, are uncontrolled).
Because both terms use the same $0$--$10$ scale, we add them with equal weights as a reporting convention.
This choice does not imply that security and utility carry equal operational value in every deployment.
We therefore use $S$ only as a summary statistic and report the malicious and benign axes separately in \autoref{tab:main_results} and \autoref{tab:benign_utility}, and jointly in the Pareto comparison of \autoref{fig:pareto}.

\section{Results}
\label{sec:results}
We report the empirical evaluation of AlcaTRAz under the protocol of \autoref{sec:experiments}, using the judge-based 0--10 scoring scheme of \autoref{sec:evaluation-metric}.
Safety and utility are read \emph{jointly} throughout, because a defense can dominate either axis alone at the expense of the other, as Llama Guard does on safety.

\subsection{Overall Reduction in Jailbreak Success Rate}
\label{sec:aggregate_reduction}

Pooling the malicious-prompt judge scores over all 33 models and 23 conditions (849\,321 scored responses per setting), the undefended distribution is heavily right-skewed and peaks at score~10 ($251\,883$ responses, $\approx 29.7~\%$); after defense the modal score shifts to~2 ($240\,880$, $\approx 28.4~\%$), so the defense redistributes the \emph{whole} distribution toward refusal rather than only clipping the most successful jailbreaks.
The mean malicious score falls from 6.65 to 4.48 ($-32.6~\%$), the share in the top band (8--10) from 49.8~\% to 26.9~\%, and the refusal share (0--2) rises from 6.1~\% to 50.1~\%.
We stress, however, that the defended distribution stays \emph{bimodal}: read as a thresholded success rate, $26.9~\%$ of responses still land in the 8--10 band and would leak in a deployment that tolerates no severe output.
AlcaTRAz therefore substantially reduces, but does not eliminate, jailbreak success, and is best deployed as one layer within a defense-in-depth strategy (\autoref{sec:limitations}).

\subsection{Results by Jailbreak Attack Type}
\label{sec:results_by_attack}
\begin{figure}[tbp]
    \centering
    \begin{subfigure}[b]{0.49\textwidth}
        \centering
        \includegraphics[width=1.02\linewidth]{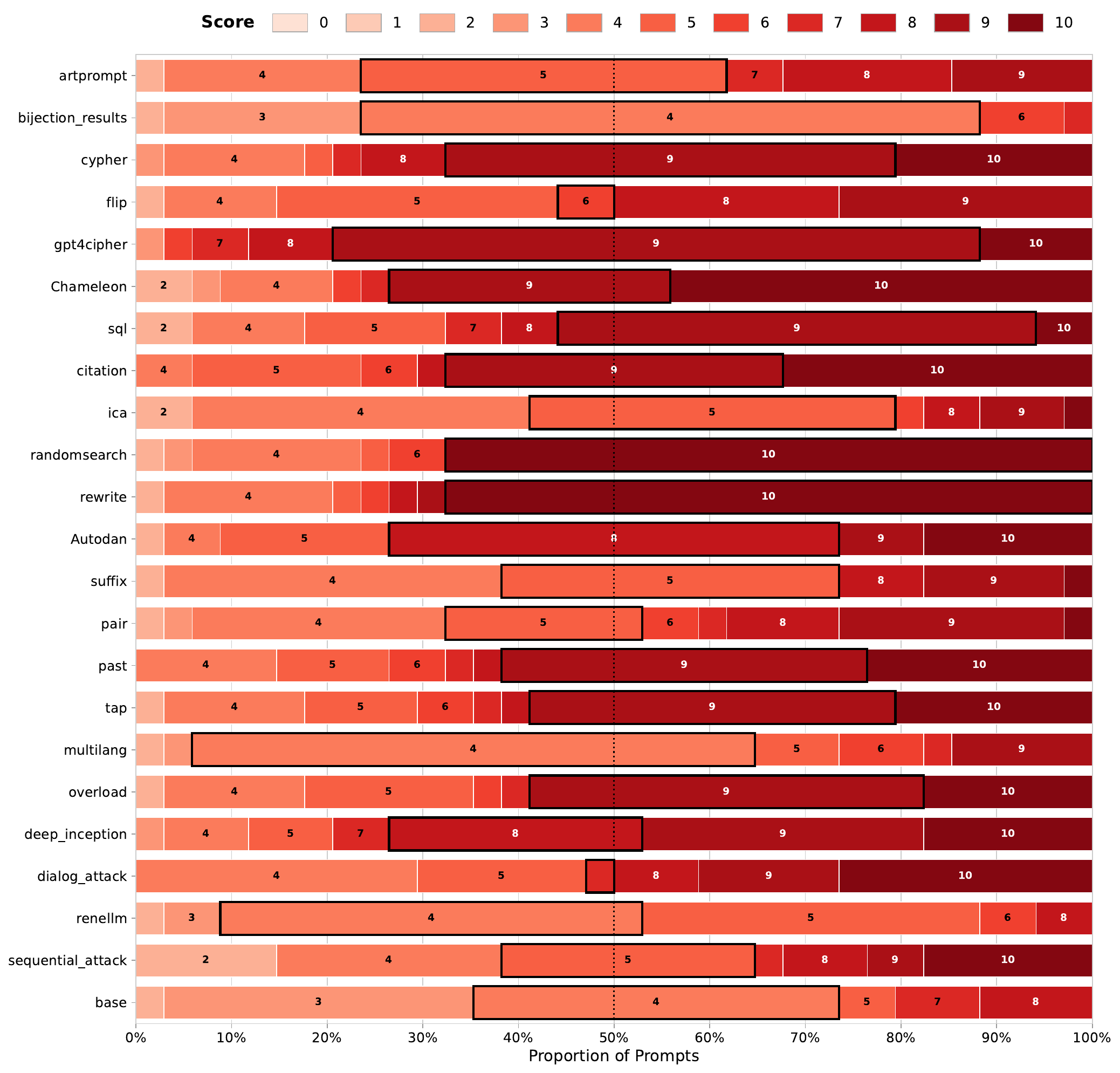}
        \caption{Before defense (undefended).}
        \label{fig:left_scores_detail}
    \end{subfigure}
    \hfill
    \begin{subfigure}[b]{0.49\textwidth}
        \centering
        \includegraphics[width=1.02\linewidth]{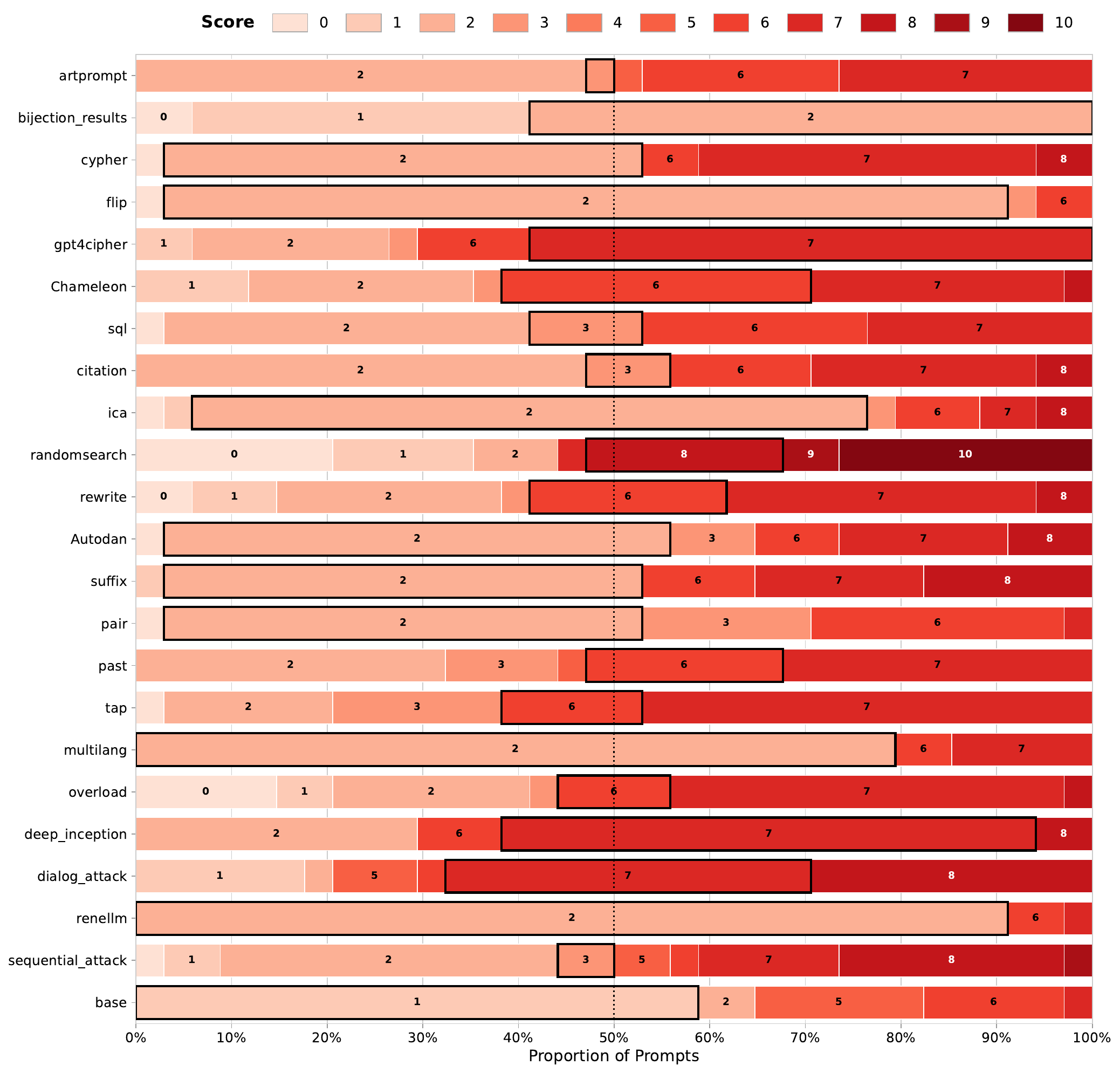}
        \caption{After applying AlcaTRAz.}
        \label{fig:right_scores_detail}
    \end{subfigure}
    \caption{Score distribution by attack type, aggregated over 33 target models. Each row shows the proportion of responses at each score level, with the median indicated by a black box. After defense (AlcaTRAz)~(b), most medians decrease, and the proportion of high scores (8--10) is substantially reduced compared to the undefended condition~(a).}
    \label{fig:attack_scores_combined}
\end{figure}

Disaggregating by attack family (\autoref{fig:attack_scores_combined}) reveals wide variation in intrinsic strength: undefended (\autoref{fig:left_scores_detail}), \texttt{rewrite}, \texttt{randomsearch}, and \texttt{Chameleon} concentrate at scores 9--10, while encoding-based attacks (\texttt{bijection}, \texttt{ica}) are broader and lower.
After defense (\autoref{fig:right_scores_detail}), the median falls by a median of 3 points and the upper tail contracts; 21 of the 23 conditions show a strict median reduction, with the largest drops on cipher- and context-based families (e.g.,\ \texttt{cypher} $9\!\to\!2$, \texttt{citation} $9\!\to\!3$).

\noindent\textbf{Attack-strength heterogeneity.}
Intrinsic strength ranges from near-harmless (\texttt{base}, \texttt{multilang}; median ${\le}4$) to saturated (\texttt{randomsearch}, \texttt{rewrite}; median~10), so aggregate success rates conflate defense quality with benchmark composition, which the per-family view avoids.
AlcaTRAz improves safety on essentially every dangerous family; the exception is optimization-based \texttt{randomsearch} (median~8 after defense), whose suffix can be re-optimized against small input changes.

\subsection{Comparison with Baseline Defenses}
\label{sec:defense_comparison}

\begin{figure}[tbp]
    \centering
    \includegraphics[width=1.0\linewidth]{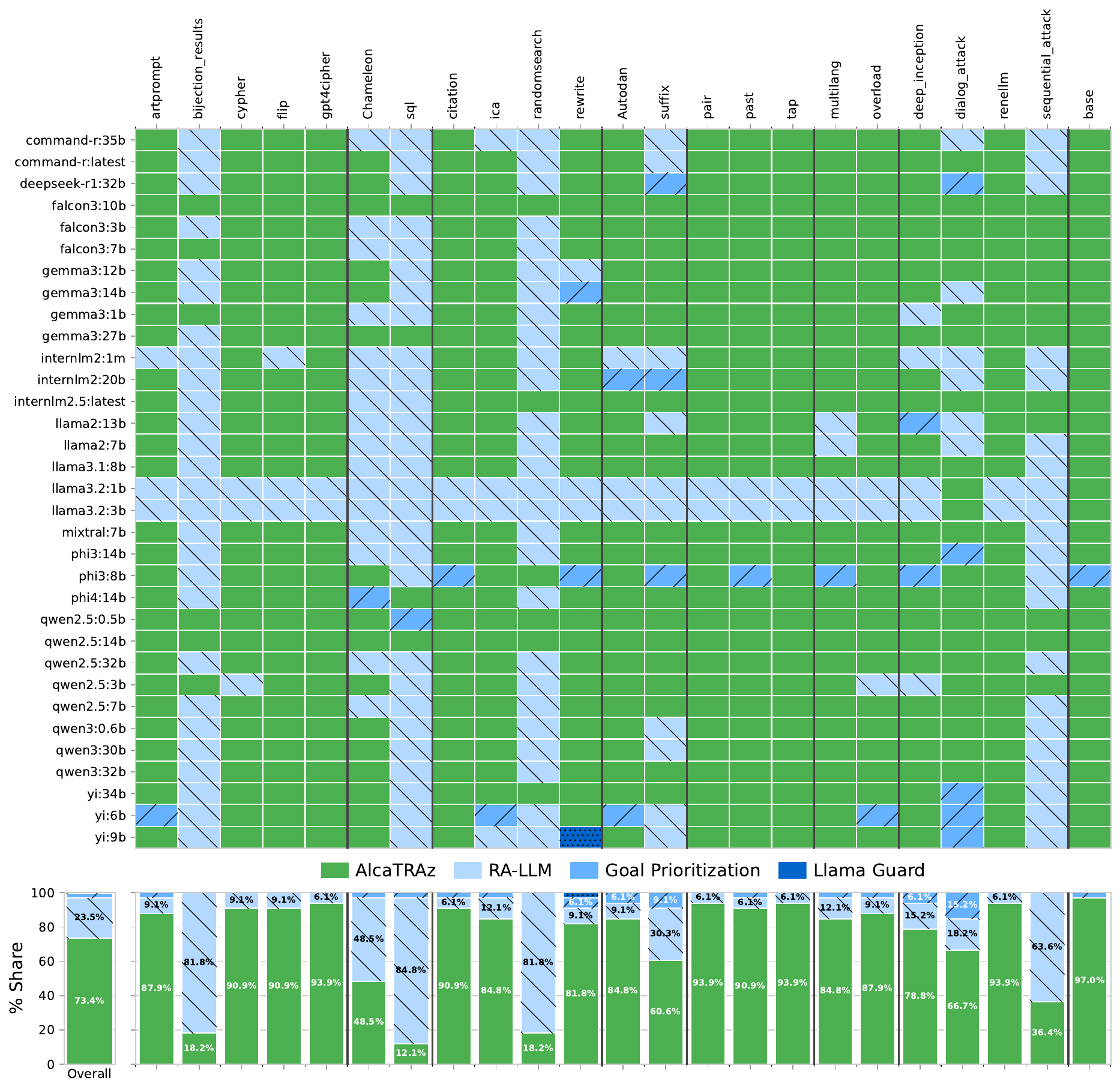}
    \caption{Best-performing defense for each target model (rows) and attack type (columns) under the composite score $S$ of~\autoref{eq:composite}. AlcaTRAz (green) wins in 73.4~\% of cells, RA-LLM (light blue) in 23.5~\%, Goal Prioritization (blue) in 3.0~\%, and Llama Guard (dark blue) in 0.1~\%. The bottom panel shows aggregate shares across all models.}
    \label{fig:defense_comparison}
\end{figure}

\autoref{fig:defense_comparison} compares AlcaTRAz with the undefended setting and with the three prompt-level baselines across all 33 evaluated models and 22 attack types.
Each cell indicates which defense achieved the best trade-off between malicious-prompt compliance and benign-prompt functionality for that model-attack combination, evaluated jointly through the composite score $S$ of \autoref{eq:composite}.
AlcaTRAz wins in 73.4~\% of all combinations, compared with 23.5~\% for RA-LLM, 3.0~\% for Goal Prioritization, and fewer than 0.1~\% for Llama Guard.
\autoref{tab:main_results} summarizes the aggregate safety statistics, and \autoref{tab:benign_utility} reports the benign-functionality statistics used in the composite score.

\begin{figure}[tbp]
    \centering
    \includegraphics[width=0.85\linewidth]{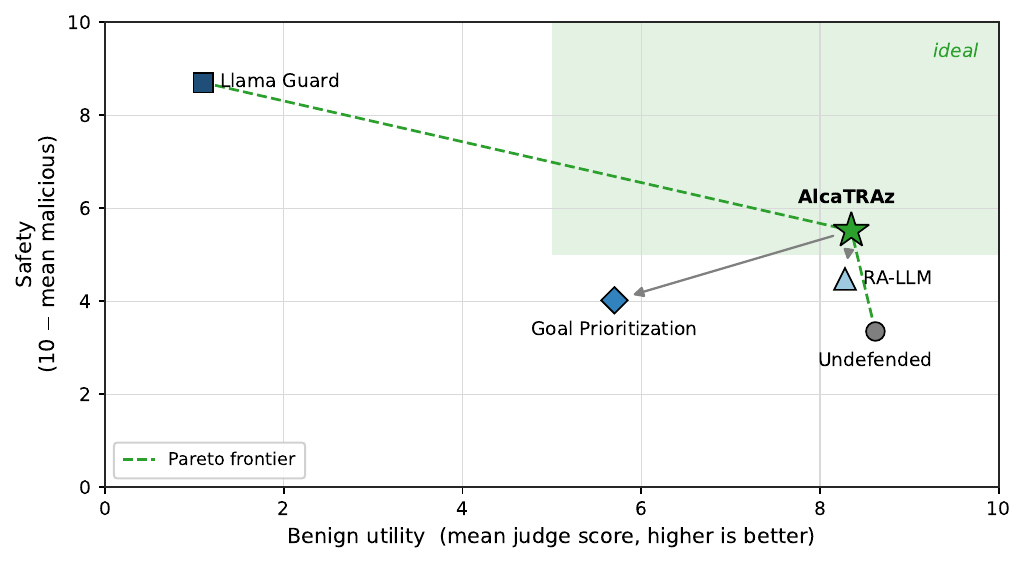}
    \caption{Trade-off between security and utility for the five conditions. Horizontal: benign utility (higher is better); vertical: safety ($10$ minus mean malicious compliance, higher is better). The green region is the ideal quadrant, safe and useful at once. AlcaTRAz Pareto-dominates the utility-preserving baselines (RA-LLM, Goal Prioritization; gray arrows), while Llama Guard reaches high safety by giving up benign utility.}
    \label{fig:pareto}
\end{figure}

\paragraph{Reading the trade-off directly.}
A natural concern is whether the edge of AlcaTRAz is an artifact of $S$, since Llama Guard attains a lower \emph{absolute} malicious score (\autoref{tab:main_results}).
\autoref{fig:pareto} reports the two axes directly and resolves this question.
Llama Guard reaches its low malicious score only by reducing benign answerability.
Among the utility-preserving baselines, AlcaTRAz \emph{Pareto-dominates} RA-LLM and Goal Prioritization, and it is the only condition in the ideal region.
The equal-weight composite score therefore formalizes a dominance relation that is already visible in the per-axis numbers.

\begin{table}[tbp]
\centering
\captionsetup{font=footnotesize}
\setlength{\tabcolsep}{4pt}
\renewcommand{\arraystretch}{1.1}
\resizebox{\columnwidth}{!}{%
\begin{tabular}{lcccc}
\toprule
\textbf{Defense} & \textbf{Avg.\ Score $\downarrow$} & \textbf{Scores 8--10 [\%] $\downarrow$} & \textbf{Scores 0--2 [\%] $\uparrow$} & \textbf{Best by $S$ [\%]} \\
\midrule
Undefended              & 6.65 & 49.8 &  6.1 & --    \\
Llama Guard             & 1.30 &  6.0 & 93.3 &  0.1  \\
RA-LLM                  & 5.52 & 52.7 & 40.2 & 23.5  \\
Goal Prioritization     & 5.98 & 54.8 & 31.9 &  3.0  \\
\textbf{AlcaTRAz}  & \textbf{4.48} & \textbf{26.9} & \textbf{50.1} & \textbf{73.4} \\
\bottomrule
\end{tabular}%
}
\vspace*{0.5em}
\caption{Aggregate safety statistics across all 33 models and 22 attack types. Arrows indicate the preferred direction. The rightmost column reports the percentage of model-attack combinations on which the defense obtains the highest composite security and functionality score $S$ of \autoref{eq:composite}.}
\label{tab:main_results}
\vspace*{-1.5em}

\end{table}

The advantage of AlcaTRAz is broad rather than localized.
For 27 of the 33 evaluated models, it achieves the best composite score across most conditions.
At the attack level, AlcaTRAz obtains the best composite score on the majority of the 33 models for 18 of the 22 attack types.
The four exceptions are \texttt{bijection}, \texttt{sql}, \texttt{randomsearch}, and \texttt{sequential\_attack}\footnote{All belong to gradient-based or optimization-heavy families in which random perturbation is a strong ad hoc baseline.}, on which the token-level dropout of RA-LLM obtains the higher composite score.
\begin{table}[tbp]
\centering
\scriptsize
\setlength{\tabcolsep}{6pt}
\renewcommand{\arraystretch}{1.1}
\begin{tabular}{lcc}
\toprule
\textbf{Defense} & \textbf{Benign Score} $\uparrow$ & \textbf{$\Delta$ vs.\ Undefended} \\
\midrule
Undefended              & 8.62 & -- \\
Llama Guard             & 1.10 & $-7.52$ \\
RA-LLM                  & 8.28 & $-0.34$ \\
Goal Prioritization     & 5.70 & $-2.92$ \\
\textbf{AlcaTRAz}  & \textbf{8.35} & \textbf{$-0.27$} \\
\bottomrule
\end{tabular}
\vspace*{0.5em}
\caption{Benign-query functionality across defenses, measured as the average judge score on 1\,120 benign Quora questions. Higher scores indicate better functionality preservation.}
\label{tab:benign_utility}
\vspace*{-2em}
\end{table}
Even on those attacks, AlcaTRAz trails RA-LLM only narrowly, while outperforming it across the remaining attack types.

\subsection{Benign-Query Functionality}
\label{sec:benign_utility}

A prompt-level defense is practical only if it preserves utility on legitimate queries.
On the 1\,120 short, single-turn Quora questions (\autoref{tab:benign_utility}), AlcaTRAz scores 8.35 versus 8.62 undefended, a drop of only 0.27 points ($\approx 3.1~\%$).
It essentially ties RA-LLM (8.28) but at far stronger safety, while Goal Prioritization loses 2.92 points, because its system prompt makes the model cautious even on benign queries.
Llama Guard falls to 1.10, thereby substantially reducing benign answerability under our implementation.
This result may depend on classification thresholds and implementation choices and is not generalized beyond our setup.
The size of the drop warrants a dedicated error analysis of the rejected prompts (the original paper~\cite{inan2023llamaguardllmbasedinputoutput} reports no utility for benign queries), which we leave for future work.

\section{Limitations}
\label{sec:limitations}
\noindent\textbf{Adaptive attackers.}
Our evaluation applies fixed attack instances and therefore covers the defense-unaware attacker (\autoref{sec:threat_model}), but neither the defense-aware attacker as a separate condition nor the rule-aware adaptive attacker that knows $\mathcal{R}^\star$ and re-optimizes directly against the defended pipeline.
This omission is common in the prompt-level defense literature, where defenses are typically tested against fixed attack sets or generic optimizers rather than against an adversary tailored to the specific defense~\cite{nasr2025attackermoves}, and we treat it as an open gap.
The omission is consequential here: adaptive attacks are known to circumvent perturbation and smoothing defenses\footnote{Mostly $>95~\%$ success against SmoothLLM~\cite{zheng2024improvedfewshot} and perplexity filtering.}, and the attack that already resists AlcaTRAz most strongly is the optimization-based \texttt{randomsearch}, precisely the family that an attacker is most likely to exploit.
We thus claim no robustness against adaptive or optimization-based attackers; establishing it would require a nested, GCG-style optimization with the defense in the loop, which we leave to future work.

\noindent\textbf{Isolating the contribution of learning.}
Our results show that a learned anchored rule is effective, but not that learning is \emph{necessary}.
The decisive missing experiment is a matched-budget comparison against random insertion, fixed/hand-designed templates, and a no-search rule tree, alongside the paraphrasing/retokenization~\cite{jain2023baselinedefensesadversarialattacks} and randomized-smoothing~\cite{robey2023smoothllm} baselines. We also ran no sensitivity analysis over $\rho$, anchor placement, $d_{\min}$, the vocabulary, and GP hyperparameters (set from preliminary runs), nor a fitness-vs-generation study of whether $G>105$ helps; these ablations are our priority.

\noindent\textbf{Evaluation approximation.}
The same judge is both the GP fitness signal and the final evaluator. Disjoint train/eval prompts prevent prompt-level overfitting but not exploitation of \emph{this} judge's idiosyncrasies, since the fitness signal is the judge itself.
Its agreement with five annotators is only near a \emph{noisy} human ceiling (Krippendorff's $\alpha=0.511$, MAE${}=2.111$), so only aggregate, distributional shifts are reliable; confirming the ranking with an independent judge and a human study remains future work.

\noindent\textbf{Scope of utility and distribution shift.}
Benign utility is measured only on short, single-turn factual queries (Quora), and some Quora items can be borderline.
The result should not be generalized to coding, structured prompts, long-form generation, multi-turn interaction, multilingual use, or dedicated over-refusal benchmarks~\cite{rottger2024xstest,cui2025orbench}, none of which are evaluated here.
The rule is optimized for a fixed attack distribution and may require periodic re-evolution; rare zero-to-nonzero transitions (the undefended model refuses, but the defended one does not) show that perturbation is not strictly monotonic with respect to safety.

\vspace*{-1em}
\section{Conclusion}

We presented AlcaTRAz, a prompt-level defense for open-weight models that learns anchored character-level perturbations through genetic programming.
Across 33 open-weight models and 22 attack families, it obtains the best joint security and functionality trade-off among the compared baselines, shifting the modal judge score from 10 (maximal-severity) to 2 (near-refusal) and winning the composite score $S$ on 73.4~\% of model-attack combinations while keeping benign utility within 0.27 points of the undefended baseline.
The improvement is real but bounded: a substantial high-severity tail remains (26.9~\% of attacked responses still score 8--10) and we do not evaluate adaptive attackers, so AlcaTRAz is best deployed as one mitigation layer within a defense-in-depth strategy rather than a standalone guarantee.
Beyond the concrete defense, these results argue for evaluating prompt-level safeguards jointly on security and utility, and they motivate further controlled study of automated search as an alternative to manually specified perturbation rules~\cite{guo2024evoprompt,fernando2024promptbreeder}.

\begin{credits}

\vspace*{-0.5em}
\subsubsection{Ethics statement.}
This work develops defensive technology for LLM safety. All experiments use public datasets, open-weight models, and previously published attacks; no real users were exposed to harmful content, and annotators were informed that the content being assessed may be harmful.

\vspace*{-0.5em}
\subsubsection{AI Declaration.}
AI tools were used only for clarity, grammar, and phrasing; all ideas, methods, experiments, and conclusions are the authors' own.

\vspace*{-0.5em}
\subsubsection{\ackname} This work was partially supported by the Brno University of Technology (internal project FIT-S-23-8151).
This research was supported by Red Hat.
\end{credits}

\bibliographystyle{splncs04}
\bibliography{bibliography}

\end{document}